          [1999/12/02 1.01 (PWD)]
\documentclass[a4paper,twocolumn]{esapub} 
\usepackage{natbib}
\input epsf.tex

\title{MODELING FLARES PRODUCED IN BLAZARS}
\author{Marek Sikora}
\author{Micha{\l} B{\l}a\.zejowski}
\affil{Nicolaus Copernicus Astronomical Center, Warsaw}
\author{Greg Madejski}
\affil{Stanford Linear Accelerator Center, Menlo Park, CA 94025, USA}
\author{Rafa{\l} Moderski}
\affil{JILA, University of Colorado, Boulder, CO }

\begin{document}


\maketitle

\begin{abstract}

We model light curves of flares observed in blazars, assuming
that they are produced by relativistic electrons/positrons injected
by forward-reverse shocks traveling down the jet
with relativistic speeds.  We approximate the radiating region as a thin
shell enclosed between the two shock fronts.  In addition 
to the conventional radiation processes such as synchrotron radiation and its
Comptonization, we take into account Comptonization of light
originating in the broad emission line region, 
and Comptonization of IR radiation from hot dust. 
We demonstrate that flares produced in the adiabatic regime decay much slower
than flares produced in the radiative regime. The lack of such a difference
between the X-ray and $\gamma$-ray flares in 3C279 strongly supports 
the synchrotron-self Compton (SSC) mechanism as responsible for X-ray production in this object. 
Finally, we discuss a possible scenario which can explain the 
`cooling' nature of the MeV peak 
in blazars by comparing the flares produced in soft $\gamma$-ray band and 
above 100 MeV. 
\end{abstract}

\section{The Model}
In our model we adopt the shock-in-jet scenario, in which individual flares
are produced by shocks formed due to velocity irregularities in the beam and
traveling down the jet with relativistic speeds. 

The model assumptions are as follows:

\begin{itemize} 
\item  non-thermal plasma producing flares is enclosed within thin shells, 
each with a radial comoving width, $\lambda'$, much smaller than  
its cross-sectional radius $a$;   

\item shells propagate down the conical jet with a constant
Lorentz factor $\Gamma$;

\item magnetic fields, carried by the beam, scale with distance as 
$B' \propto 1/r$;

\item  both magnetic field intensity and  particle distribution are
uniform across the shell;

\item relativistic electrons/positrons are injected into the shell
within a finite distance range $\Delta r_{inj}$ and with injection
starting at $r_0$, at a rate parameterized by
$Q = K \gamma^{-p}$ for $\gamma_m < \gamma < \gamma_{max}$, 
and  $Q \propto \gamma^{-1}$ for $\gamma <\gamma_m$, 
where $\gamma$ is the random Lorentz factor of an electron/positron;

\item radiative energy losses of relativistic electrons/positrons are dominated
by three processes: external-radiation Comptonization (ERC):
\begin{equation} \left(d\gamma \over dt'\right)_{ERC} = - {16 \sigma_T \over 9 m_e c } 
(u_{BEL} + u_{IR}) \Gamma^2 \gamma^2  , 
\end{equation}

 synchrotron
radiation :
\begin{equation} \left(d\gamma \over dt'\right)_{SYN} = - {4\sigma_T \over 3 m_e c } u_B' \, 
\gamma^2 , 
\end{equation}

and Comptonization of synchrotron radiation :
\begin{equation} \left(d\gamma \over dt'\right)_{SSC} = - {4\sigma_T \over 3 m_e c } u_S'
\, \gamma^2 , 
\end{equation}
where $u_B' = B'^2/8\pi$ is the magnetic field energy density, $u_S'$
is the energy density of the synchrotron radiation field, $u_{BEL}$ 
$\simeq$ $({\partial L_{BEL}/ \partial \ln r}) / 4 \pi \, r^2\, c$
is the energy density of the broad emission line field  
at the actual distance of 
a source/shell propagating downstream a jet, 
$u_{IR} \simeq \xi_{IR} \, 4 \sigma_{SB} \, T^4/c$ is the energy
density of the infrared radiation field, $\xi_{IR}$ is the fraction of 
the central radiation reprocessed into near infrared by dust, 
and $T$ is the temperature of dust.

\item the observer is located at an angle $\theta_{obs} =1/\Gamma$ from the jet axis.

\item evolution of electrons is described by a continuity equation (Moderski et al. 2000; B{\l}a\.zejowski et al. 2000) 
\begin{equation} {\partial N_{\gamma} \over \partial r} = - {\partial \over \partial\gamma} \left(N_{\gamma} {d\gamma \over dr}\right) + { Q\over c\beta \Gamma} ,
\label{eq:equation1}
\end{equation}
where the rate of electron/positron energy losses is
\begin{equation} {d\gamma \over dr} = 
{1 \over \beta c \Gamma} \left(d\gamma \over dt'\right)_{rad}- 
{\gamma \over r} , \label{eq2} 
\label{eq:equation2}
\end{equation}
$\beta = \sqrt {\Gamma^2 -1}/\Gamma$, and $dr = \beta c \Gamma dt'$.  
The second term on the right-hand side of 
eq.~(\ref{eq2}) represents the adiabatic energy losses.

\end{itemize}

\section{Adiabatic vs. Radiative losses}
Here we compare the time profiles of 
flares produced by electrons radiating their energy in adiabatic 
($ \gamma \ll \gamma_c$) and radiative ($ \gamma \gg \gamma_c$) regimes,
where $\gamma_c$ is the energy of electrons for which the radiative 
cooling time scale is equal to the injection time scale.
Presented flares are produced by the ERC process and all 
computations are made under assumption that this process strongly dominates
the entire radiative losses.  A comparison of two profiles in Fig. 1 
illustrates that adiabatic flares decay much more slowly than 
radiative ones.  This property of flares produced in the two regimes 
can be used to verify X-ray production mechanisms in blazars 
(see next section).
 
\begin{figure}[h]
\centerline{\epsfysize=8.0truecm\epsfbox{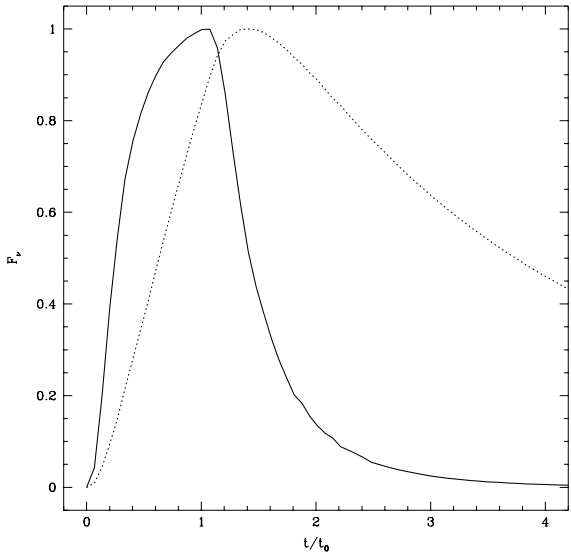}}
\caption{Adiabatic (dotted line) vs. radiative (solid line) flares. Note that $t_0 = r_0/(1+\beta)\beta c \Gamma^2 \simeq r_0/2 c \Gamma^2$.}
\end{figure}

\section{Confrontation with observations - 3C 279}
3C 279 is among the best monitored blazars over a broad range of 
energies.  In this section we use the simultaneous
multi-wavelength observations carried out in February 1996 
(Wehrle et al. 1998) to infer about the radiative processes operating
in this object.  The most interesting result of this campaign was 
the very close correspondence between the X-ray and $\gamma$-ray light 
curves during the outburst. There was no evidence of any significant 
time lag between the peaks;  furthermore, both curves decayed at the 
same rate. The above facts suggest the following scenario: X-rays 
are produced via SSC the process and $\gamma$-rays by ERC (see Fig. 2 
for details).  We can definitely exclude a model where both energy
bands are produced by the ERC process alone.  This is due to the fact 
that the production of X-rays by ERC involves low energy electrons, 
which radiate in the adiabatic regime. As was shown in previous
section, adiabatic flares decay much more slowly than radiative ones. 
Thus, the observed $\gamma$-ray -- X-ray correlation excludes the 
production of X-rays by the ERC process, unless the decay rate is 
determined by the decelaration of the shell and/or a change in 
its direction of motion. 
 
\begin{figure}[h]
\centerline{\epsfysize=8.5truecm\epsfbox{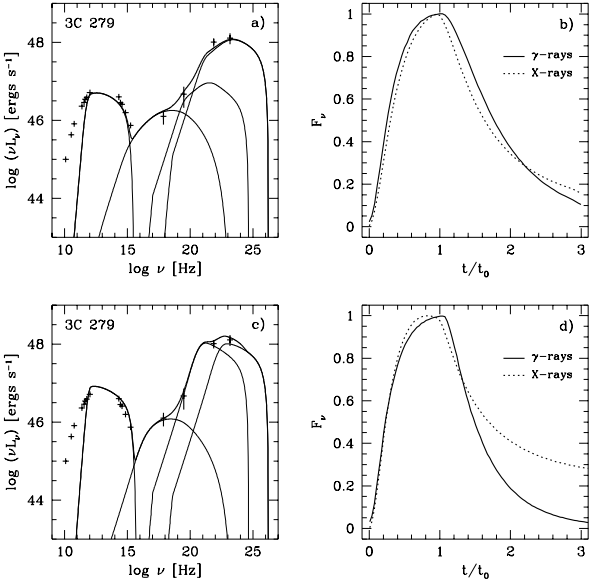}}
\caption{Modeling  the Feb-96 event in 3C 279. 
Left panels: two model fits to the time averaged 
outburst spectrum. All observational data are  simultaneous
and taken from Wehrle et al. (1998). Right panels: model $\gamma$-ray
flare, with the flux integrated over energies $> 100$ MeV, vs. model X-ray flare, 
with the flux taken at 2 keV. The upper panels are for the model with
parameters: $\Gamma=20$; $r_0 = 6.0 \times 10^{17}$ cm;  
$\Delta r_{inj}/r_0 =1$; $Q=const$, with 
$\gamma_m = 27$, $\gamma_{max}=6.5 \times 10^3$ and $p=2.4$, and electron
injection luminosity
$L_e  = 1.6 \times 10^{44}$ ergs s$^{-1}$;
$B' = 0.53 (r_0/r)$ Gauss; and energy densities of the ambient diffuse radiation field
s, 
$u_{BEL}= 4.9 \times 10^{-4} (r_0/r)^2$ ergs cm$^{-3}$ and
$u_{IR}= 1.0 \times 10^{-5}$ ergs cm$^{-3}$.
The lower panels are for the model with parameters: $\Gamma=20$; 
$r_0 = 7.0 \times 10^{17}$ cm;  
$\Delta r_{inj}/r_0 =1$; $Q=const$, with 
$\gamma_m = 150$, $\gamma_{max}=6.5 \times 10^3$ and $p=2.4$, and
$L_e  = 0.6 \times 10^{44}$ ergs s$^{-1}$;
$B' = 0.81 (r_0/r)$ Gauss; and energy densities of the ambient diffuse radiation field
s, 
$u_{BEL}= 5.9 \times 10^{-4} (r_0/r)^2$ ergs cm$^{-3}$ and
$u_{IR}= 2.0 \times 10^{-4}$ ergs cm$^{-3}$. }
\end{figure}

\section{The MeV Break-predictions vs. observations}
In previous sections it was assumed that the break 
in the electron injection function, $\gamma_m$, 
is located at lower energies than the break caused by inefficient radiative
cooling of electrons with $\gamma < \gamma_c$.
With this assumption, the characteristic break
of the high energy spectra  in the 1--30 MeV range
is related to the break of the electron energy distribution at $\gamma_c$. 
As was discussed by Sikora (1997),
such an interpretation of the MeV break is consistent with the observed 
flare time scales. Assuming that the distance of flare production
is of the order $r_{fl} \sim c t_{fl} \Gamma^2$, one can find that flares lasting $
\sim$days are produced at distances $10^{17}-10^{18}$cm, and that at such distances
 ERC radiation is  
inefficient at $h\nu < 1-30$ MeV. 

\begin{figure}[h]
\centerline{\epsfysize=8.5truecm\epsfbox{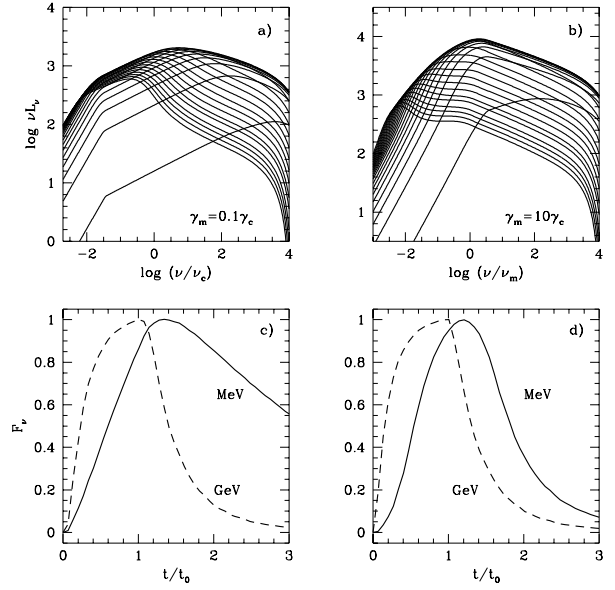}}
\caption{Variability around the ERC peak.
Left panels are for the peak produced by electrons with energies
$\gamma \sim \gamma_c$, while right panels are for the peak determined
by electrons with $\gamma \sim \gamma_m$. The spectra are shown with a time
step $\delta t/t_0 = 0.14$ and are followed up to $t = 3t_0$. 
The flares are shown at frequencies: (c) $\nu = 0.1 \nu_c$ (solid line)
and $\nu = 100 \nu_c$ (dashed line), where 
$\nu_c = (16/9) \Gamma^2 \gamma_c^2 \nu_{IR}$, ($\nu_{IR}$ is the frequency of the
Comptonized photon); 
(d) $\nu = 0.1 \nu_m$ (solid line)
and $\nu = 100 \nu_m$ (dashed line), where 
$\nu_m = (16/9) \Gamma^2 \gamma_m^2 \nu_{IR}$. 
$(d\gamma/dt')_{ERC}$ is assumed to be distance independent.}
\end{figure}

However, the time resolution of recent high-energy experiments is not 
good enough to reject the possibility that the observed
flares are superpositions of several shorter-lasting flares, which are 
produced much closer to the central engine than the $\sim 1-$day 
variability time scales may suggest. Due to the stronger magnetic 
and external radiation fields at smaller distances, the value of 
$\gamma_c$ would be lower and the break would appear at 
$h\nu \ll 1$ MeV. In order to explain the observed MeV breaks, 
one would then have to assume that this break is related to the 
break in the electron injection function.  There are two ways to 
distinguish between these possibilities, either by studying spectral 
slopes, or by studying flare profiles, both below and 
above the MeV break. The first approach is based on the fact that
for $\gamma_m < \gamma_c$, the slope of the radiation spectrum produced
by electrons with $\gamma_m \ll \gamma \ll \gamma_c$ should be
harder by $\delta \alpha = 0.5$ than the slope of the radiation
spectrum produced by electrons with $\gamma \gg \gamma_c$.
But in the case $\gamma_c < \gamma_m$, the slope of 
the  radiation spectrum  produced by electrons with 
$\gamma_c \ll \gamma \ll \gamma_m$ is predicted to be $\alpha = 0.5$, 
independent of the spectral slope produced at $\nu \gg \nu(\gamma_m)$ 
(see, e.g., Sari, Piran \& Narayan 1998).  The second approach, 
proposed in this paper, uses the fact  that adiabatic flares decay
much more slowly than radiative ones (see Fig. 2).
Thus, in  the case $\gamma_m < \gamma_c$, the flares observed 
in OVV/HP quasars at sub-MeV energies should decay much more slowly 
than in the case $\gamma_c < \gamma_m$. Our predictions are presented 
for the model, for which $u_{BEL} =const$. One can see that the 
higher the energy, the more rapidly the flare peaks (Fig. 3 -- lower 
panels).  Please note that this numerical result is qualitatively 
consistent with observations presented by Zhang et al. 2000 (these 
proceedings). These authors present simultaneous observations of
flaring blazar PKS 1622-297 showing the pronounced time lag between 
flare detected by EGRET ($E > 100$ MeV) and COMPTEL (10-30 MeV).  
Observations indicate a time delay of about 4 days between the two bands.

\end{document}